\documentclass[12pt,epsf,epsfig,psfig]{article}
\usepackage{graphicx}
\usepackage{epsfig}
\def\J{$J/\psi$}
\def\j{J/\psi}

\def\P{$\psi'$}
\def\p{\psi'}

\def\be{\begin{equation}}
\def\ee{\end{equation}}

\def\lsim{\raise0.3ex\hbox{$<$\kern-0.75em\raise-1.1ex\hbox{$\sim$}}}
\def\gsim{\raise0.3ex\hbox{$>$\kern-0.75em\raise-1.1ex\hbox{$\sim$}}}

\def\NP{{ Nucl.\ Phys.\ }}
\def\PL{{ Phys.\ Lett.\ }}
\def\PR{{ Phys.\ Rev.\ }}

\def\PRL{{ Phys.\ Rev.\ Lett.\ }}

\def\ZP{{ Z.\ Phys.\ }}
\def\EP{{ Eur.\ Phys.\ J.}}

\title{
\vskip1.5cm
{\bf Parton Percolation in Nuclear
Collisions}}

\author{~~\\
{\bf Helmut Satz} \\
~~\\
{\normalsize Fakult\"at f\"ur Physik, Universit\"at Bielefeld}\\
{\normalsize D-33501 Bielefeld, Germany}\\
{\normalsize and}\\
{\normalsize Departamento de F{\' \i}sica, Instituto Superior T\'ecnico}\\
{\normalsize P-1049 001 Lisboa, Portugal}}

\begin{document}

\date{}

\maketitle

\thispagestyle{empty}



\begin{abstract}

An essential prerequisite for quark-gluon plasma production in nuclear
collisions is cross-talk between the partons from different nucleons in
the colliding nuclei. The initial density of partons is determined by
the parton distribution functions obtained from deep inelastic
lepton-hadron scattering and by the nuclear geometry; it increases with
increasing $A$ and/or $\sqrt s$. In the transverse collision plane, this
results in clusters of overlapping partons, and at some critical density,
the cluster size suddenly reaches the size of the system. The onset of
large-scale cross-talk through color connection thus occurs as geometric
critical behavior. Percolation theory speci\-fies the details of this
transition, which leads to the formation of a condensate of deconfined
partons. Given sufficient time, this condensate could eventually
thermalize. However, already the onset of parton condensation in the
initial state, without subsequent thermalization, leads to a number of
interesting observable consequences.

\end{abstract}

\vskip 1cm
\newpage

\section{Initial State Conditions}

\bigskip

Statistical QCD predicts that with increasing temperature, strongly
interacting matter will undergo a transition from a hadronic phase to a
plasma of deconfined quarks and gluons. In the hadronic state, the
chiral symmetry of the QCD Lagrangian (for massless quarks) is
spontaneously broken; in the quark-gluon plasma, it is restored. These
predictions are the result of finite temperature lattice QCD studies,
and for the calculations it is crucial that they deal with a thermal
medium, i.e., with equilibrium thermodynamics.

\medskip

\begin{figure}[htb]
\centering
\resizebox{0.6\textwidth}{!}{%
\includegraphics*{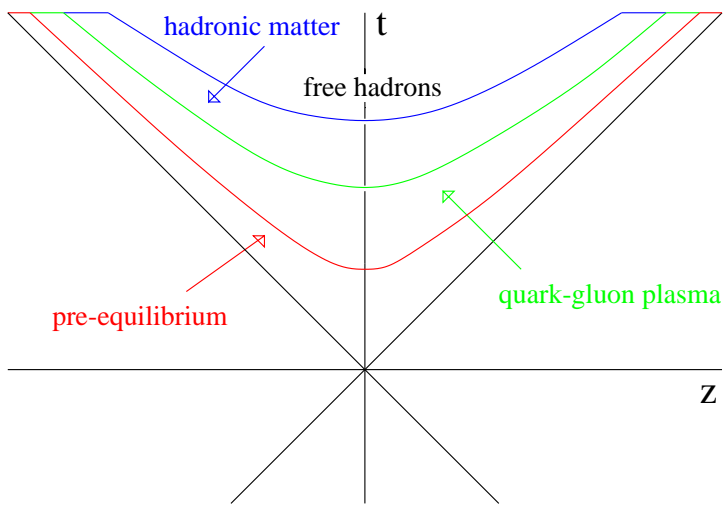}}
\vglue3mm
\caption{The expected evolution of a nuclear collision}
\label{evo}
\end{figure}

\begin{figure}[ht!]
\centering
\begin{tabular}{ccccc}
\resizebox{0.38\textwidth}{!}{%
\includegraphics*{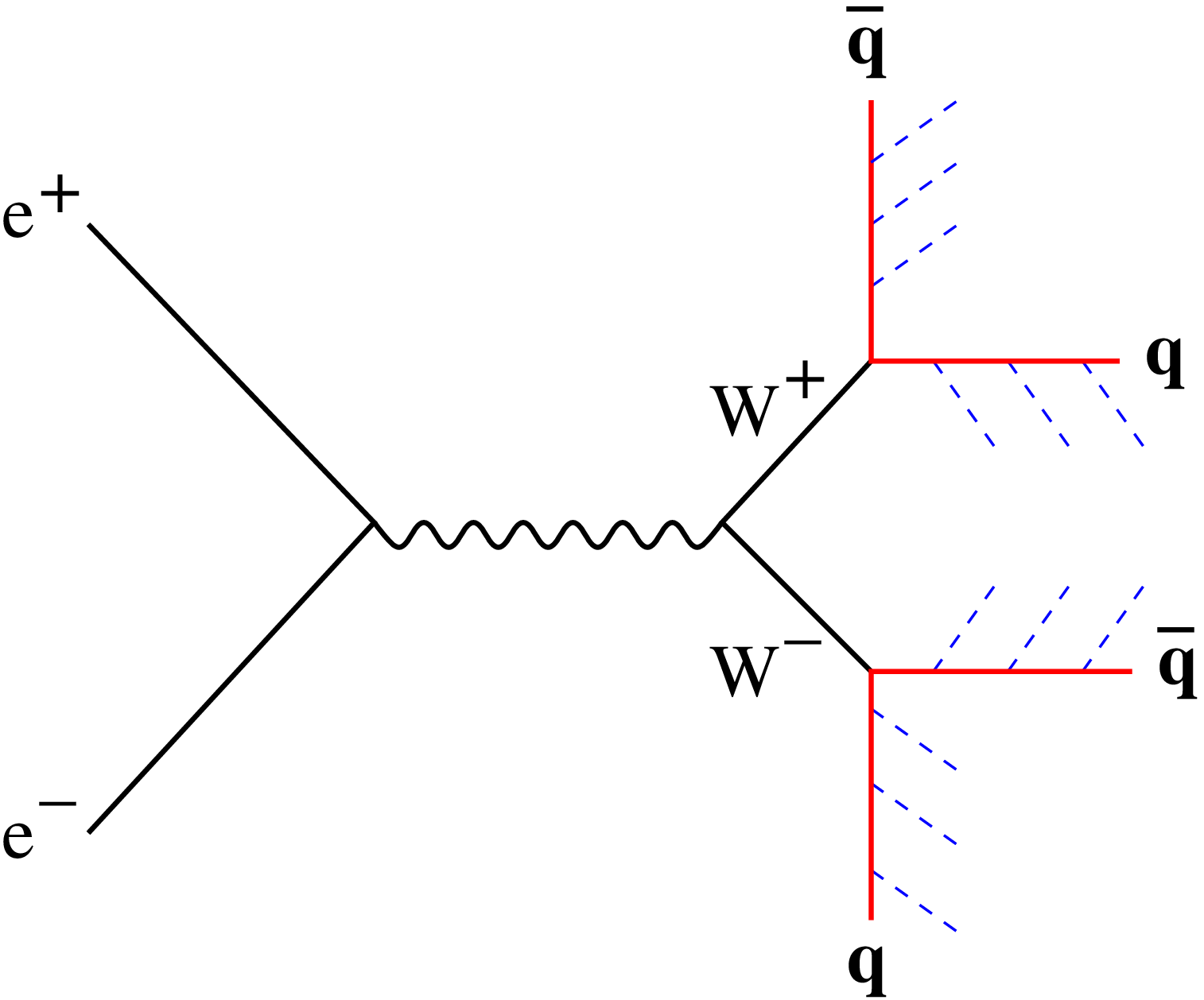}}
\hskip 0.9cm
\resizebox{0.38\textwidth}{!}{%
\includegraphics*{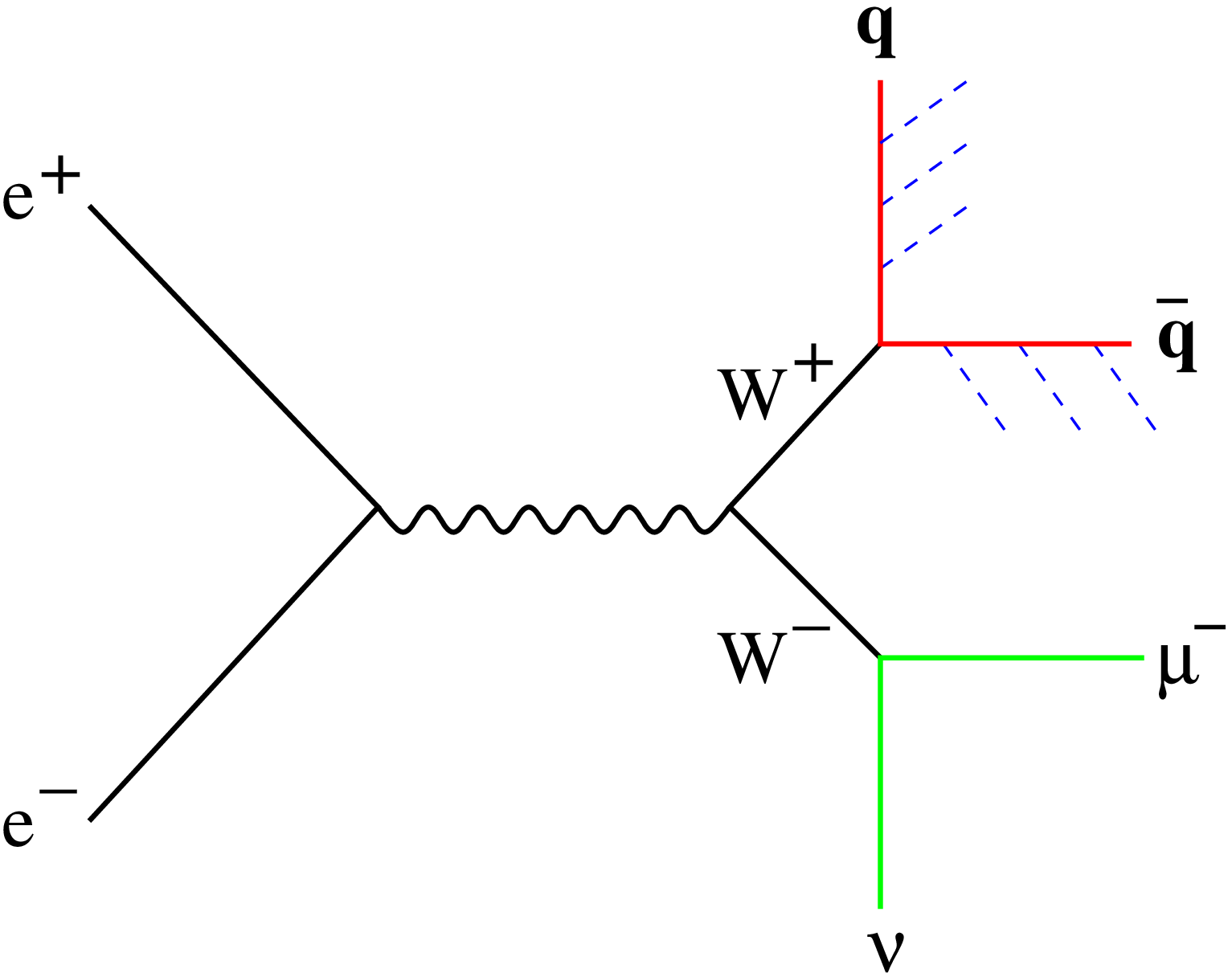}}
\end{tabular}
\caption{Four-jet and two-jet decays of $W^+W^-$ pairs in $e^+e^-$
annihilation at LEP.}
\label{WW}
\end{figure}

The initial state of two colliding nuclei is clearly a non-equilibrium
configuration. The canonical view of its evolution is schematically
illustrated in Fig.\ \ref{evo}. After the collision, there is a short
pre-equilibrium stage, in which the primary partons of the colliding
nuclei interact, multiply and then thermalize to form a quark-gluon
plasma. This QGP then expands, cools and hadronizes. A prerequisite for
the equilibration process is evidently that the partons originating
from different nucleons form a large-scale interconnected system. If
there is no ``cross talk" between partons from different nucleons,
thermalization is not possible. The problem of color connection has
been studied in hadron production through $W^+W^-$ decay at LEP. The
$W$'s are produced essentially at rest in the annihilation of an
energetic $e^+e^-$ pair (see Fig.\ \ref{WW}), and it is possible to
compare the reaction in which both $W$'s produce hadronic jets to that
in which one decays leptonically. If there is cross talk between the
decay quarks of one $W$ with those from the other, the multiplicity
of the four-jet decay is predicted to be less than twice that in the
two-jet decay \cite{KKG}. The data show no such reduction, suggesting
that the decay quarks from different $W$'s don't communicate \cite{LEP}.

\medskip

It is therefore necessary to determine under what conditions the initial
state parton configurations can lead to color connection, and more
specifically, if variations of the initial state can lead to
a transition from disconnected to connected partonic clusters.
The results of such a study of the pre-equilibrium state in nuclear
collisions do not depend on the subsequent evolution and thus in
particular not require any kind of thermalization.

\medskip

The structural problem underlying the transition from disconnected to
connected systems of many components is a very general one, ranging
from clustering in spin systems to the formation of galaxies. The
formalism is given by percolation theory, which describes geometric
critical behavior \cite{stauffer}. We shall return to the basic idea
a little later on.

\section{Partons in Nuclei}

\begin{figure}[htb]
\vglue4mm
\centering
\resizebox{0.7\textwidth}{!}{%
\includegraphics*{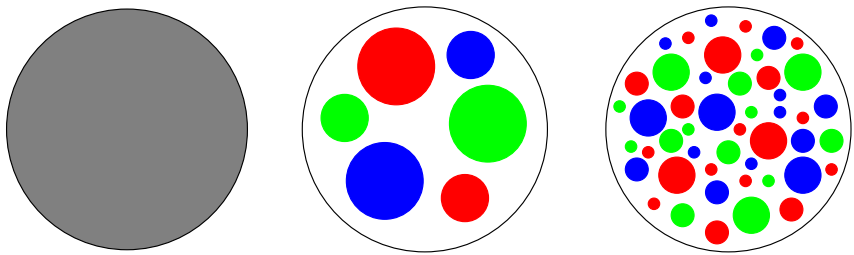}}
\vglue3mm
\caption{The structure of an incoming nucleon seen (left to right)
for increasing resolution}
\label{bees}
\vglue4mm
\end{figure}

If you look at a fast nucleon coming at you, what do you see? The answer
depends on who's looking. Another nucleon or a pion sees a disc of
radius $r \simeq$ 1 fm and a certain greyness. A hard photon, with a
resolution scale $Q^{-1} \ll 1$ fm, sees a swarm of partons. How many
there are depends on the resolution scale: given a finer scale, you can
see smaller partons, and there are more the harder you look 
(Fig.\ \ref{bees}). The partons in a nucleon have a transverse size $r_T$
determined by their intrinsic transverse momentum $k_T$, with $r_T
\simeq 1/k_T$. The scale $Q^{-1}$ specifies the minimum $k_T^{-1}$
resolved, so the probing photon sees all partons in the range $0 \leq
k_T \leq Q$.

\medskip

\medskip

The momentum $p$ of the incoming nucleon is distributed among the
partons; a parton of momentum $k$ carries the fraction $x=k/p$. In deep
inelastic scattering experiments, the distribution of partons in a
nucleon is determined for given $x$ and $Q$. Denoting the gluon content
by $g(x,Q)$, that of quarks and antiquarks by $q(x,Q)$ and $\bar
q(x,Q)$, respectively, we have the overall momentum conservation sum
rule
\be
\int_0^1 dx ~x ~\{g(x,Q) + \sum_i [q_i(x,Q) + \bar q_i(x,Q)] \} =1,
\label{1}
\ee
where $i$ counts the number of quark flavors.

\medskip

The number of partons in a nucleon at rapidity $y$, as seen by a photon
of scale $Q$, is thus given by
\be
{dN \over dy} =
x ~\{g(x,Q) + \sum_i [q_i(x,Q) + \bar q_i(x,Q)] \},
\label{2}
\ee
where $x=(k_0+k_L)/(p_0+p_L)$ in terms of the parton and nucleon
energies $k_0,~p_0$ and longitudinal momenta $k_L,~p_L$. Since the
scale $Q$ specifies the maximum $k_T$ resolved, $dN/dy$ in Eq.\
(\ref{2}) gives us the total number of partons in the range $0 \leq k_T
\leq Q$.

\medskip

In a minimum bias nucleon-nucleon collision, the transverse parton size
itself determines the resolution: it sets the scale at which partons
`probe each other' in the colliding nucleons, so that here the highest
relevant $k_T$ fixes $Q$. Since at $y=0$, the fractional momentum is
$x=k_T/\sqrt s$, Eq.\ (\ref{2}) provides at given $\sqrt s$ the total
number of partons of transverse momenta up to $Q$.

\medskip

\begin{figure}[htb]
\centerline{\psfig{file=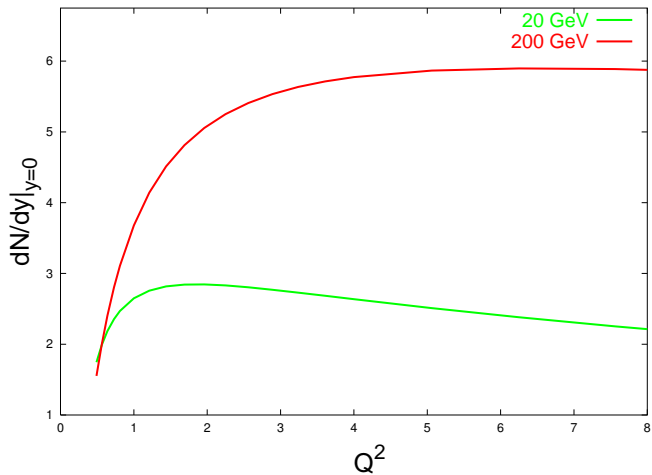,width=6cm,angle= -90}}
\vskip 0.8cm
\caption{Parton density at central rapidity as function of the
resolution scale $Q$ at $\sqrt s=20$ and 200 GeV, using PDF GRV94DIS.}
\label{GRV}
\end{figure}

As mentioned, the quark and gluon distributions in a nucleon are
determined from deep inelastic scattering data. In Fig.\ \ref{GRV}, we
show the resulting variation of $(dN/dy)_{y=0}$ as function of
$Q^2 \simeq <k_T^2>$ for two values of the c.m.s.\ energy, $\sqrt s =
20$ GeV (SPS) and 200 GeV (RHIC), using the GRV94DIS parametrization
\cite{GRV}, which is particularly suitable for our kinematic range.
It is evident that with increasing $Q^2$, more and more partons of
decreasing size come into play, so that $(dN/dy)_{y=0}$ increases
strongly with $Q$. It is also clear that at higher collision energy,
there are more partons -- at RHIC two times more than at SPS.  The
slow decrease at large $Q^2$ is due to kinematic constraints: at fixed
$\sqrt s$, increasing $Q^2$ means increasing $x$ and hence decreasing
gluon or sea quark density.

\medskip

\section{Partons in Nuclear Collisions}

\begin{figure}[htb]
\vspace*{0.5cm}
\hspace*{1.9cm}
\mbox{
\hskip1cm
\epsfig{file=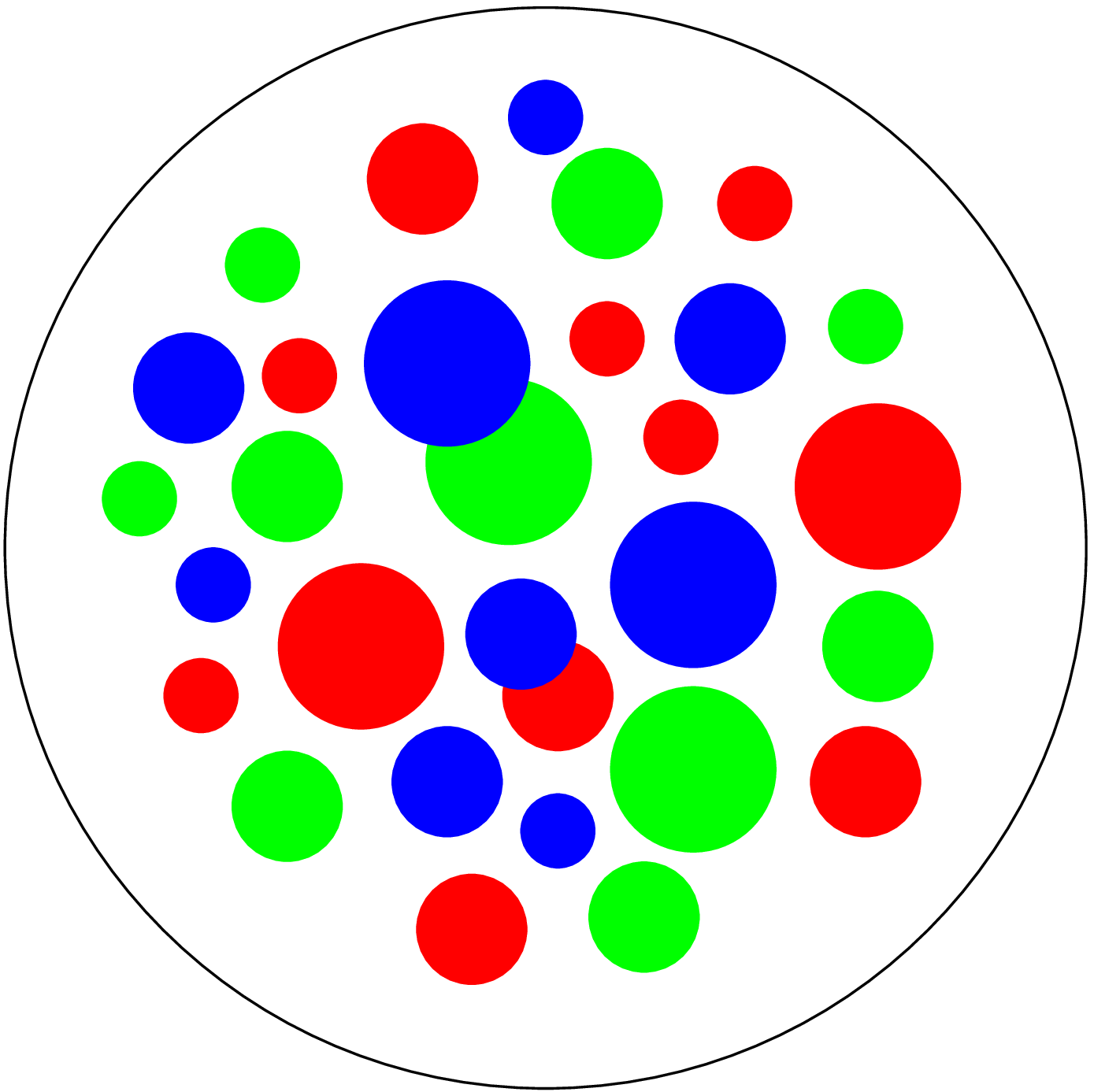,width=4cm}
\hskip2cm
\epsfig{file=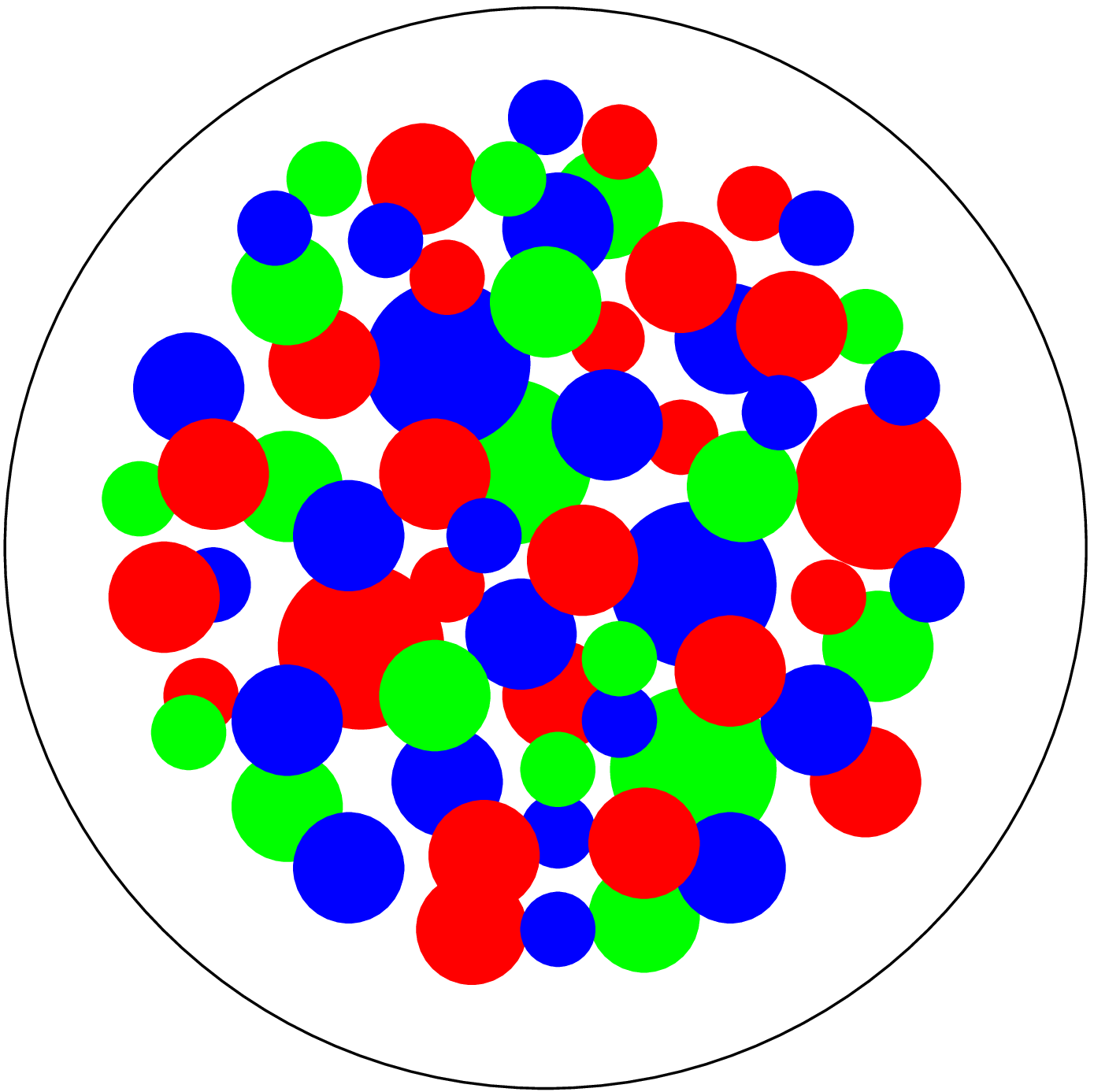,width=4cm}}
\vskip0.5cm
\caption{Partonic cluster structure in the transverse collision plane
at low (left) and high (right) parton density}
\label{discs}
\end{figure}

Consider now the collision of two heavy nuclei at high energy, as seen
in the overall center of mass. The Lorentz-contraction in the
longitudinal direction makes it a collision of two thin discs, so that
in the transverse plane, the parton density increases with $A$.
The partons from different nucleon begin to overlap and form clusters:
see Fig.\ \ref{discs}. How does the cluster size grow with parton
density, and when does it reach the dimension of the total transverse
collision area? These are precisely the questions addressed by
percolation theory, so that here we make a small interlude.

\subsection{Percolation Theory}

\begin{figure}[htb]
\vspace*{-0.5cm}
\hspace*{0.5cm}
\centerline{\psfig{file=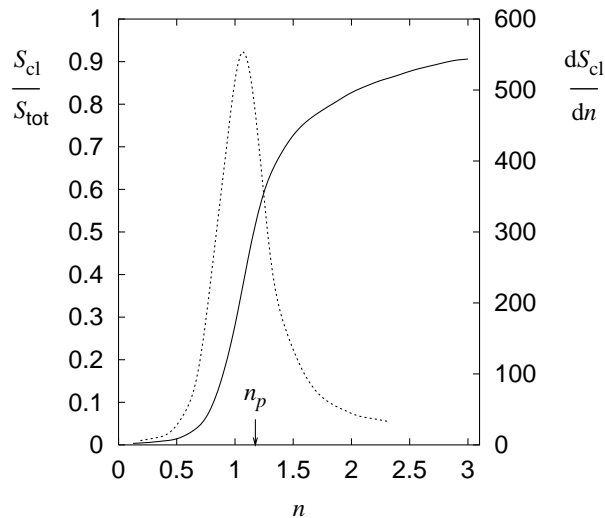,height=70mm}}
\vspace*{0.1cm}
\caption{The fractional cluster size and its derivative as function of
the parton density $n$}
\label{cluster}
\end{figure}

Consider placing $N$ small circular discs (`partons') of radius $r$
onto a large circular manifold (`the transverse nuclear plane') of
radius $R\gg r$; the small discs may overlap. With increasing parton
density $n \equiv N/\pi R^2$, this overlap will lead to more and larger
connected partonic clusters. The striking feature of this phenomenon is
that the average cluster size $S_{\rm cl}$ does not grow as some power
of $n$; instead, it increases very suddenly from very small to very
large values (see Fig.\ \ref{cluster}). This suggests some kind of
geometric critical behavior, and in fact in the `thermodynamic limit'
$N\to \infty,~R\to \infty$, the cluster size diverges at a critical
threshold value $n_c$ of the density $n$,
\be
S_{\rm cl} \sim (n_c - n)^{-\gamma},
\label{gamma}
\ee
as $n \to n_c$ from below. This appearence of infinite clusters at
$n=n_c$ is defined as percolation: the size of the cluster reaches the
size of the system. The divergence is governed by the critical exponent
$\gamma=43/18$, determined analytically, while the threshold
\be
n_c = {1.128 \over \pi r^2}
\label{n-c}
\ee
is obtained numerically or through analytical approximation \cite{isi}.
Hence we obtain in the limit of large systems
\be
{N\over \pi R^2 } = {1.128 \over \pi r^2}
\label{perco-cond}
\ee
as percolation condition.
Note that because of parton overlap, the manifold is at percolation not
totally covered by discs, even though the overall disc area slightly
exceeds that of the manifold: $N~\pi r^2 = 1.128~ \pi R^2$. In fact,
one can show that at $n=n_c$, the fraction
\be
1 - \exp\{-1.128\} \simeq 0.68
\label{fraction}
\ee
of the area $\pi R^2$ is covered by partonic discs.

\medskip

\medskip

Before we return to the study of nuclear collisions, we want to
comment briefly on the relation between percolation and thermal
phase transitions \cite{stauffer,attig}. Some thermal critical
behavior, such as the magnetization transition for ferromagnetic
spin systems, can be equivalently formulated as percolation.
However, percolation seems to be a more general phenomenon and in
particular can occur even when the partition function is analytic,
i.e., when there is no thermal critical behavior. A specific example
of this is the Ising model in a non-vanishing external field, which
has a percolation transition even though there is no magnetization
transition.

\subsection{Parton Percolation}

The results of the previous subsection tell us that in nuclear
collisions there is indeed, as function of parton density, a sudden
onset of large-scale color connection. There is a critical density at
which the partons form one large cluster, losing their independent
existence and their relation to the parent nucleons. Parton percolation
\cite{Torino,DFPS}
is thus the onset of color deconfinement and although it is a necessary
prerequisite for any subsequent QGP formation, it does not require or
imply any kind of parton thermalization.

\medskip

To obtain quantitative predictions, we have to specify the relevant
scales. The partonic size is through the uncertainty relation determined
by its average transverse momentum,
\be
\pi r^2 \simeq {\pi \over <k_T^2>},
\label{k-T}
\ee
and as mentioned, for a given resolution scale, $<k_T^2> \simeq Q^2$.
Given the parton density in a nucleon, we now have to specify the
density in a nucleus-nucleus collision. At SPS energy ($\sqrt s \simeq
20$ GeV), the wounded nucleon model appears to work quite well, so that
we have in a central $A-A$ collision
\be
\left( {dN \over dy} \right)_{y=0}^{AA} \simeq 2~A \left( {dN\over dy}
\right)_{y=0}.
\label{N-AA}
\ee
It is clear, however, that at higher energies, collision dependent
contributions will come into play \cite{K-N}.

\medskip

\begin{figure}[h]
\hskip 0.5cm
\centerline{\small{$\sqrt s = 20$ GeV}}
\vskip -0.4cm
\centerline{\psfig{file=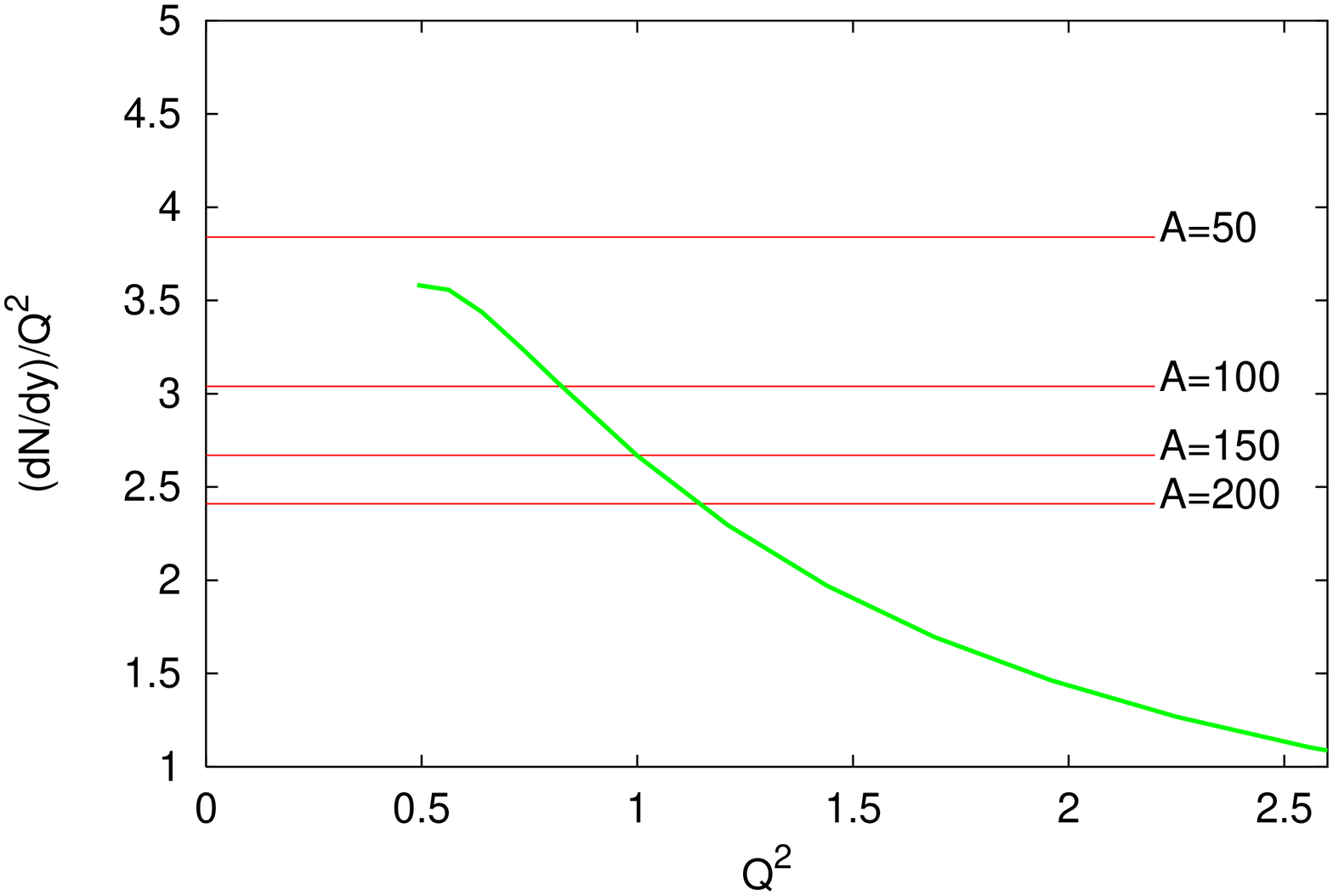,width=6cm,angle=-90}}
\vskip 0.2cm
\hskip 0.5cm
\centerline{\small{$\sqrt s = 200$ GeV}}
\vskip -0.4cm
\centerline{\psfig{file=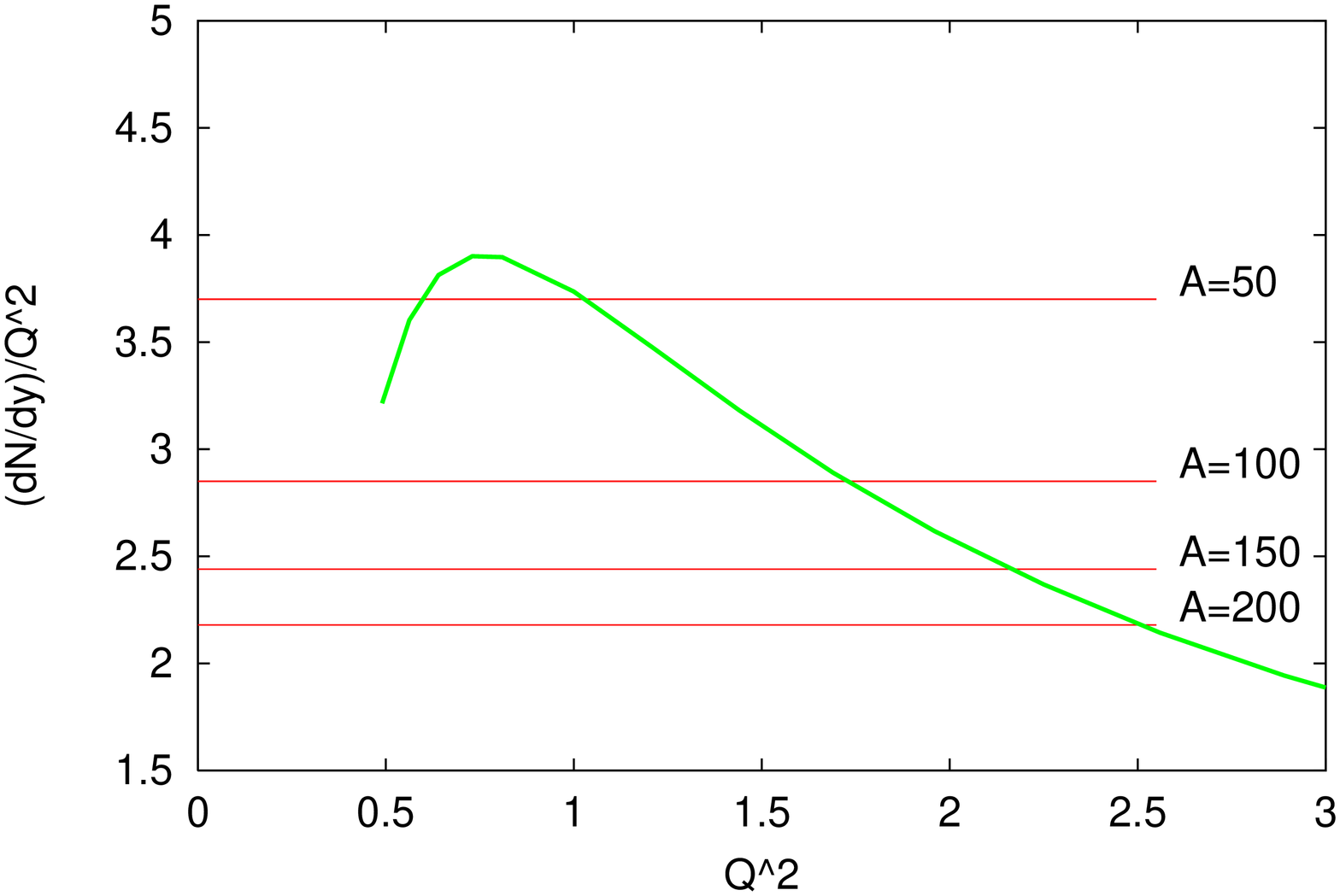,width=6cm,angle=-90}}
\caption{Percolation values of $Q^2$ for different $A$ and
collision energies.}
\label{AA}
\end{figure}

For central $A-A$ collisions, we thus obtain the percolation condition
\be
{2A \over \pi A^{2/3}} \left( {dN\over dy} \right)_{y=0} = {1.128 \over
\pi Q^{-2}}
\label{perco-AA}
\ee
in terms of $A$, the resolution scale $Q$ and the nucleonic parton
density obtained in deep inelastic scattering. Let us separate the basic
parton contributions from the nuclear dependence and rewrite eq.\
(\ref{perco-AA}) as
\be
{1 \over Q^2} \left( {dN\over dy} \right)_{y=0} = {1.128 \over
2A^{1/3}}.
\label{perco}
\ee
When the l.h.s.\ of this equation, determined by P.D.F.'s, becomes
equal to the r.h.s., fixed by nuclear size, we have the onset of
percolation. In Fig.\ \ref{AA}, the results are shown for typical SPS
and RHIC energies. We thus find that for $\sqrt s =20$ GeV, there is
percolation for $A \gsim 60$, while for ($\sqrt s =200$ GeV) it sets in
somewhat earlier, for $A \gsim 40$.

\medskip

\begin{figure}[h]
\hskip 0.5cm
\centerline{\small{$\sqrt s = 20$ GeV, Pb-Pb}}
\vskip -0.4cm
\centerline{\psfig{file=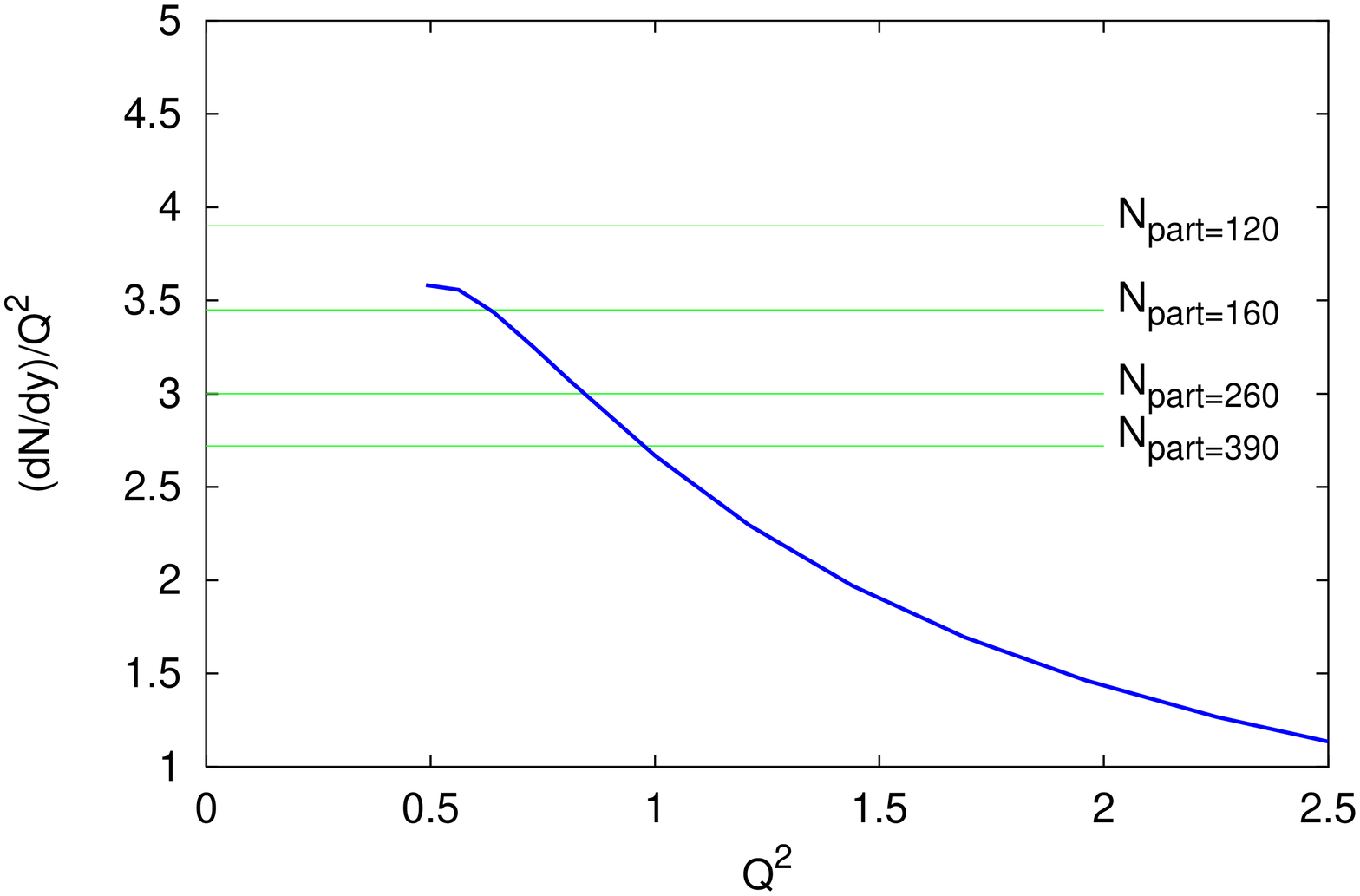,width=6cm,angle=-90}}
\vskip 0.2cm
\hskip 0.5cm
\centerline{\small{$\sqrt s = 200$ GeV, Au-Au}}
\vskip -0.4cm
\centerline{\psfig{file=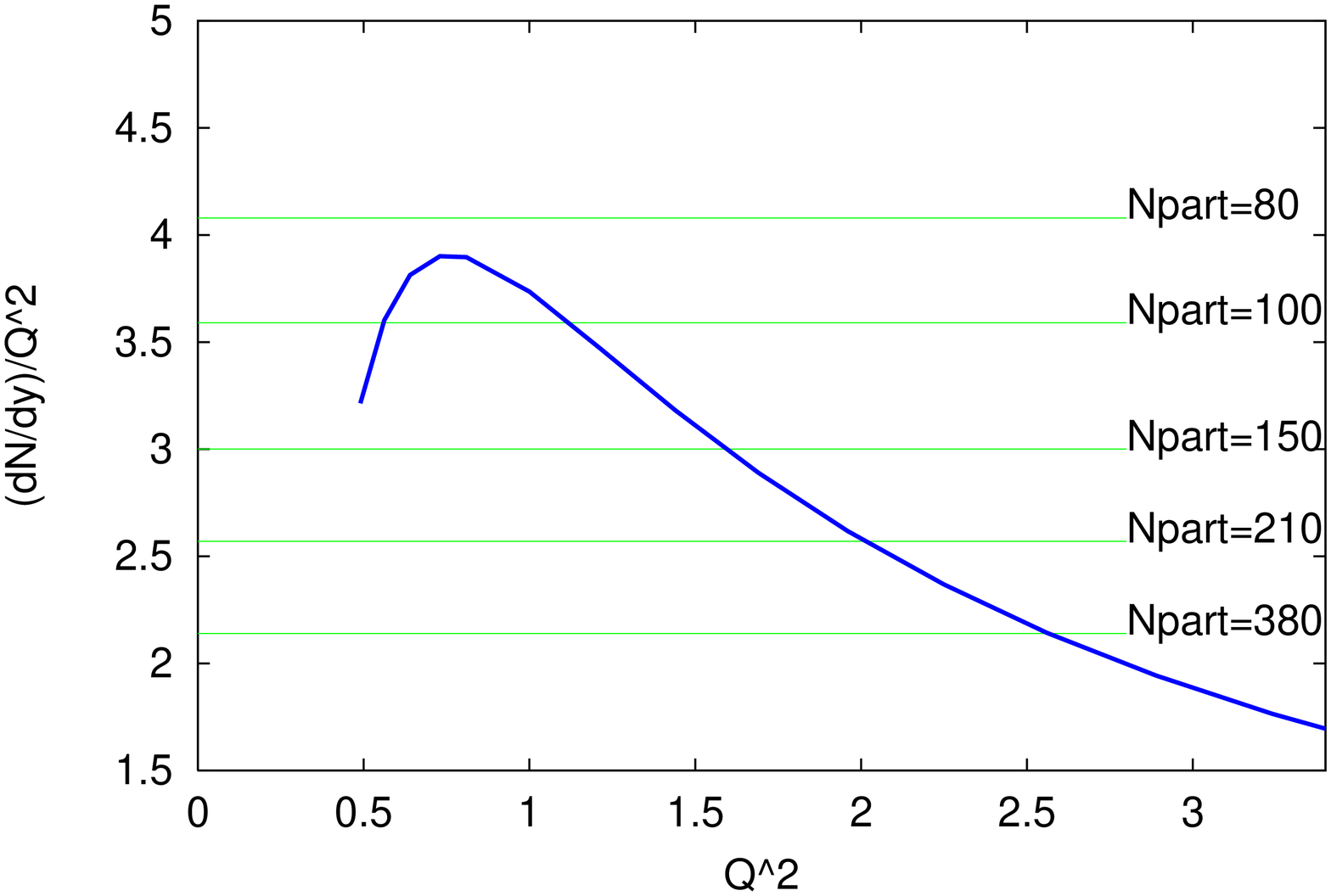,width=6cm,angle=-90}}
\caption{Percolation values of $Q^2$ for different centralities and
collision energies.}
\label{N-part}
\end{figure}

In the case of non-central $A-A$ collisions, the manifold disc size $\pi
R^2$ has to be replaced by the actual transverse overlap area at the
given impact parameter. This overlap area can be determined in a Glauber
study, using Woods-Saxon nuclear profiles \cite{KLNS}, and the resulting
counterpart of eq.\ (\ref{perco}) then leads to Figs.\ \ref{N-part}.
Here the number $N_{\rm part}$ of wounded or participant nucleons is
used to specify the centrality of the collision, since this quantity is
directly measurable. For $Pb-Pb$ collisions at $\sqrt s = 20$ GeV,
this leads to an onset of percolation at $N_{\rm part}
\simeq 150$ (corresponding to an impact parameter $b \simeq 6$ fm),
while for $Au-Au$ at $\sqrt s = 200$ GeV, $N_{\rm part} \simeq 80$ (with
$b \simeq 10$ fm) is the threshold.

\medskip

\begin{figure}[h]
\hskip 0.5cm
\centerline{\small{$\sqrt s = 20$ GeV, Pb-Pb}}
\vskip -0.4cm
\centerline{\psfig{file=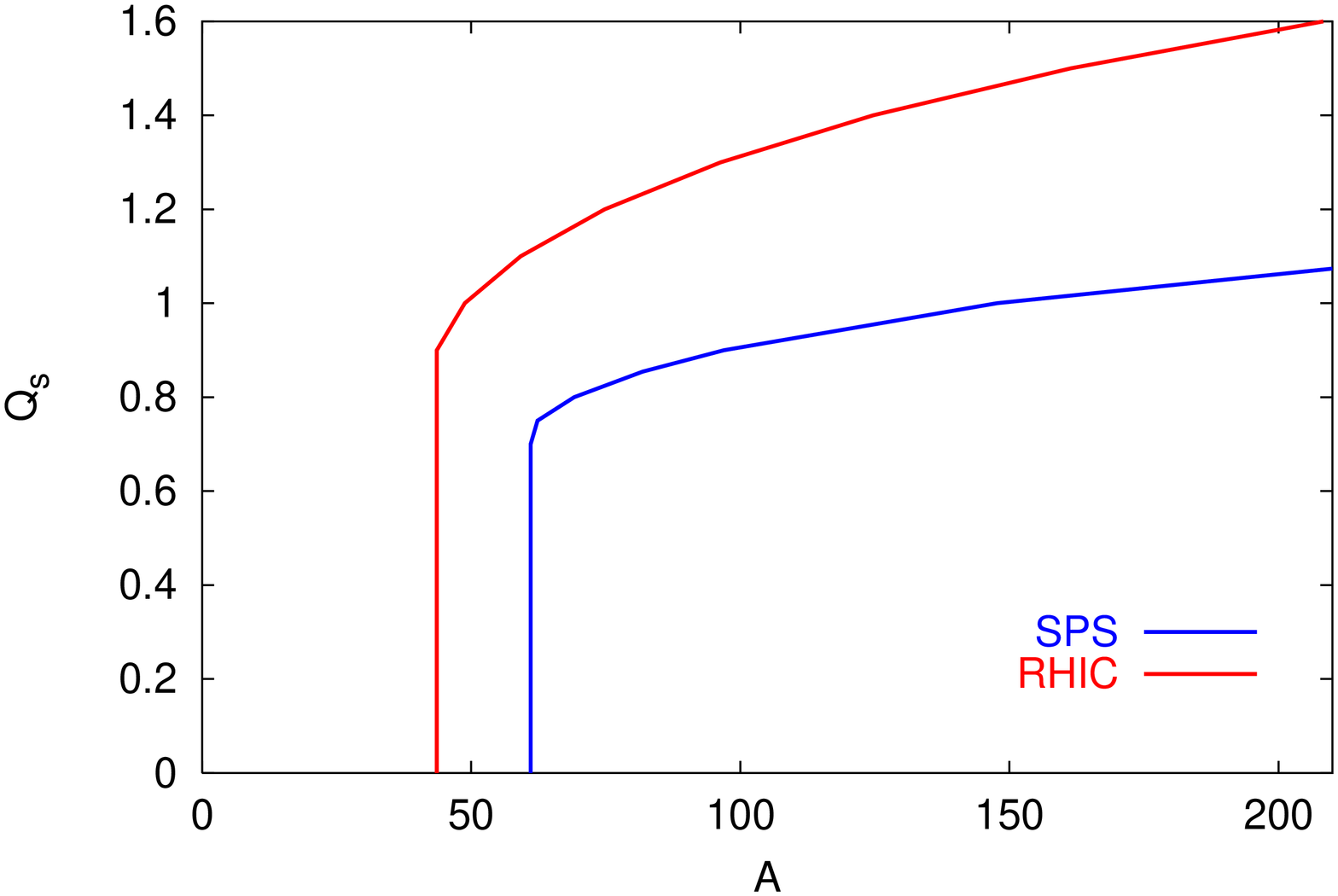,width=6cm,angle=-90}}
\vskip 0.2cm
\hskip 0.5cm
\centerline{\small{$\sqrt s = 200$ GeV, Au-Au}}
\vskip -0.4cm
\centerline{\psfig{file=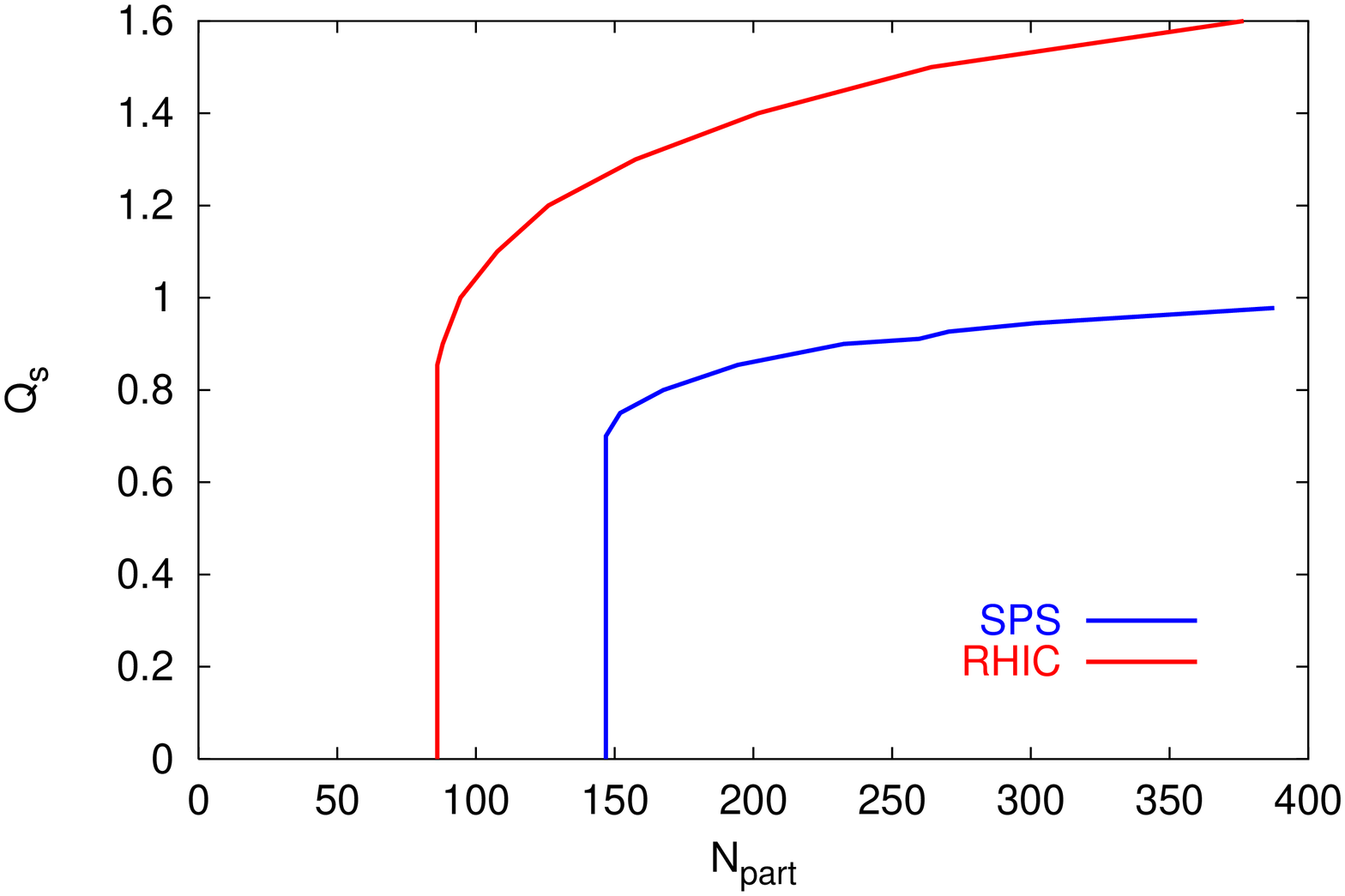,width=6cm,angle=-90}}
\caption{Percolation scale $Q_s$ vs.\ $A$ for $\sqrt s =$ 20 and
200 GeV.}
\label{Q-s-A}
\end{figure}

Beyond the percolation point, we then have a parton condensate,
containing interacting and hence color-connected partons of all scales
$k_T \leq Q$. The percolation point thus specifies the onset of color
deconfinement; it says nothing about any subsequent thermalization.
If there is eventual thermalization, the partonic momentum $k_T$ will
be related to the temperature $T$; hence the resolution scale $Q$, which
determines the range of $k_T$, is in some sense a precursor of $T$. It
is thus of interest to check how the percolation value $Q_s$ is related
to the effective nuclear size and to $\sqrt s$. This is illustrated in
Fig.\ \ref{Q-s-A} for the $A$-dependence at two energies studied here.
In the same figure, we also show the corresponding results as function
of centrality at $A=200$. It is seen that bigger $\sqrt s$, larger $A$
or more central collisions lead to a `hotter' parton condensate, in the
mentioned pre-thermal sense.

\section{Observable Consequences}

We have seen that in nuclear collisions, the parton structure of nucleons
leads to critical behavior in the form of parton percolation in the
transverse collision plane. This critical behavior is independent
of any subsequent thermalization; it is determined by the initial
collision conditions in the pre-equilibrium stage.

\medskip

The parton condensate which is formed through percolation is closely
related to the color glass condensate \cite{CGC} studied for nuclear
collisions in the limit of large $A$ and/or $\sqrt s$;
the different approaches
simply focus on different aspects. In percolation studies, the central
topic is the onset of parton condensation in terms of geometric
critical behavior. In contrast, the color glass condensate describes
the features of the high density limit for the pre-equilibrium partonic
medium, in particular in terms of classical fields.

\medskip

The occurrence of geometric critical behavior in the pre-equilibrium
stage of nuclear collisions can lead to observable consequences even
if there never is any subsequent thermalization. We note here three
possible effects of this kind: charmonium suppression, strangeness
enhancement, and energy-independent hadronic source radii.

\subsection{Charmonium Suppression}

\begin{figure}[h]
\hskip 0.5cm
\centerline{\small{A-A, with A=200}}
\vskip -0.4cm
\centerline{\psfig{file=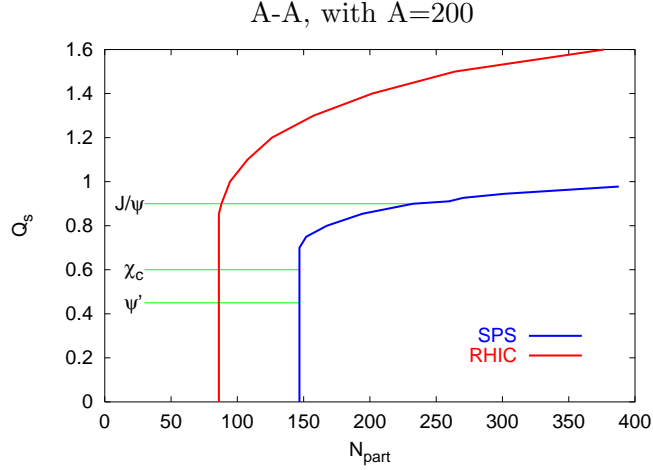,width=6cm,angle=-90}}
\caption{Charmonium dissociation as function of centrality}
\label{supp}
\end{figure}

Charmonium states are formed very early in nuclear collisions, with
a typical \J~formation time of some 0.2 - 0.3,fm obtained from binding
energy or radius. This is also the time needed for the formation of the
parton condensate, as determined by $Q_s^{-1}$. The \J~thus finds
itself in the non-equilibrium medium provided by the parton condensate.
The typical scale of the charmonium state thus has to be compared to
the intrinsic scale $Q_s$ of the parton condensate. If the latter is
indeed the precursor of temperature, it is also a precursor form of
the screening mass. We will therefore assume a charmonium state $i$ of
radius $r_i$ to be dissociated when $Q_s > r_i$. With
\be
r_{\j} \simeq (0.9~{\rm GeV})^{-1},
r_{\chi} \simeq (0.6~{\rm GeV})^{-1}
r_{\p} \simeq (0.45~{\rm GeV}^{-1},
\label{radii}\ee
for the different subthreshold charmonium states, we then obtain the
suppression values of $Q_s$ shown in Fig. \ref{supp}.

\medskip

To see what these different suppression points imply for \J~production
in nuclear collisions, we recall that in nucleon-nucleon collisions,
\J's are produced in part through feed-down. Only about 60\% of the
observed \J's are directly produced $1S$ states; of the remainder, about
30\% come from $\chi_c$ and about 10\% from \P~decay. Since these decays
occur very late in the collision evolution, the parton condensate `sees'
and hence suppresses the different charmonium states in the given
fractions. From Fig.\ \ref{supp} we thus expect that the survival
probability of \J~s in $Pb-Pb$ collisions at the SPS will show a first
anomalous suppression step at about $N_{\rm part} \simeq 150$, since
at this point the production fraction from $\chi_c$ and \P~decays
is removed. A second drop would be expected around $N_{\rm part} \simeq
250$, since now the directly produced \J's are dissociated.

\medskip

The crucial consequence is the predicted two-step suppression pattern;
to obtain reliable numerical values clearly requires the inclusion of
more details. In particular, the nuclear geometry, resulting in
percolating and non-percolating (`hot' vs.\ `cold') regions in the
transverse plane, has to be taken into account correctly in its
effect on the suppression. A two-step suppression pattern
was first obtained for \J~production in a quark-gluon plasma, since
the different charmonium radii lead to dissociation for different screening
radii. It now appears that in fact already pre-equilibrium parton
condensation leads to a similar result, so that such a pattern, if
observed, implies deconfinement but not the formation of a thermalized
quark-gluon plasma.

\medskip

\begin{figure}[h]
\hskip 0.5cm
\centerline{\psfig{file=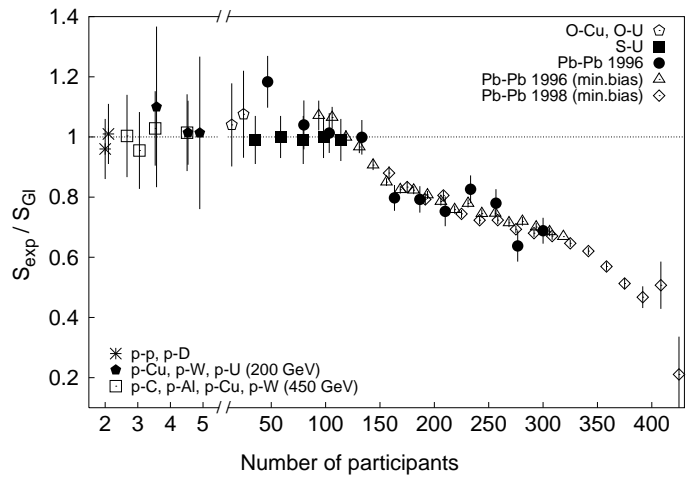,width=10cm}}
\caption{The measured \J~survival probility as function of
centrality \cite{NA50}}
\label{NA50}
\end{figure}

In this connection we note also that the onset points for charmonium
dissociation through parton condensation agree fairly well with the
`steps' seen in the measured \J~survival probability (see Fig.\
\ref{NA50}). In view of the mentioned theoretical uncertainties and
also because of possible kinematic suppression effects for very
central collisions \cite{B-O}, this agreement should so far
be considered in a more qualitative way. However, the observed
threshold value for the
onset of anomalous suppression measured in terms of the energy density
is by more than a factor two above the energy density at deconfinement.
This seems to support pre-equilibrium deconfinement through parton
condensation as underlying mechanism.

\subsection{Strangeness Enhancement}

\begin{figure}[h]
\centerline{\psfig{file=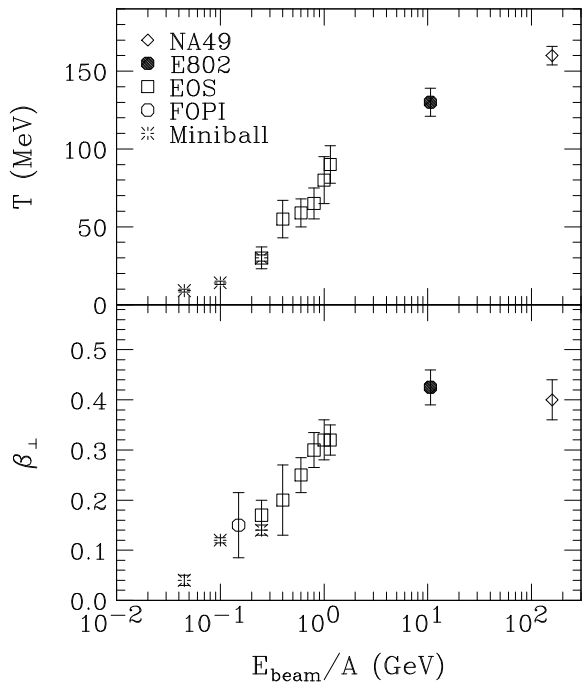,width=9cm,angle=90}}
\caption{The freeze-out temperature obtained from hadron resonance 
abundances in nuclear collisions at different energies (top), and the 
transverse expansion velocity obtained from the corresponding
hadronic $p_T$ spectra   
(bottom), compiled in \cite{Pavel}.}
\label{pavel}
\end{figure}

The relative abundance of the hadrons produced in high energy
collisions, from nucleon-nucleon to heavy ion interactions, is
well accounted for by an ideal gas of hadronic resonances \cite{BC}
The temperature of this gas converges at high energies to $T_H \simeq
150 - 180$ MeV, i.e., to the confinement temperature obtained in lattice QCD
for low baryon density (see Fig.\ \ref{pavel} for a compilation from
nucleus-nucleus collisions \cite{Pavel}).
The baryochemical potential depends on the baryon number content
of the initial state, decreasing from values around $\mu_B \simeq 0.5$
GeV in heavy ion collisions at the AGS to near zero for
$p-p/p-\bar p$ and RHIC heavy ion data.

\medskip

\medskip

The only `flaw' in this picture is that in the elementary
$p-p/p-\bar p$ collisions one observes reduced strangeness
production: the relative abundance of strange hadrons
is reduced by a factor dependent only on its content of strange
quarks/antiquarks. This strangeness suppression disappears for
high energy heavy ion collisions.

\medskip

The dependence of strangeness abundance on the density of interacting
hadrons can be accounted for if strangeness conservation is assumed to hold 
locally \cite{KR}. In a thermal medium of temperature $T$ and overall
volume $V$, the relative phase space weight (Boltzmann factor) for the
presence of a strange particle of mass $m$ is given by $\exp\{-m/T\}$.
If, however, the medium contains only one such strange particle, and if
strangeness conservation is taken to hold locally within some
correlation volume $V_0 \ll V$, then the correct Boltzmann factor
should be $\exp\{-2m/T\}$, since the simultaneous presence of the
strange particle and its antiparticle requires the expenditure of
energy $2m$. The crucial point here is the assumed
local nature of strangeness conservation: the factor $\exp\{-m/T\}$
would be correct if the single strange particle could be compensated
by a `far away' antiparticle somewhere else in the overall volume $V$.
It is only the requirement of zero strangeness within $V_0$ that leads
to the enhanced suppression. Since a given strange particle and its
antiparticle in a really {\it ideal} gas will eventually separate beyond
$V_0$, the requirement of local strangeness conservation appearently
implies the introduction of a dynamical (i.e., non-ideal) correlation.

\medskip

\begin{figure}[h]
\centerline{\psfig{file=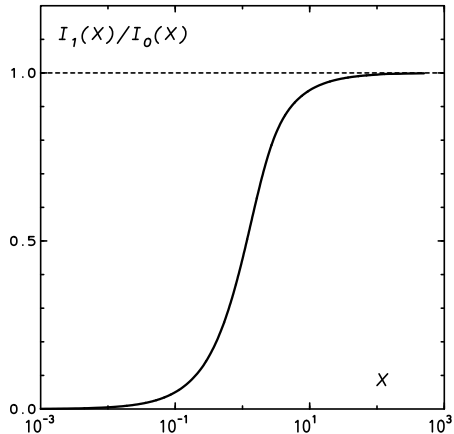,width=6cm,angle=-180}}
\vskip -1cm
\caption{The transition from local to global strangeness conservation}
\label{local}
\end{figure}

To illustrate the effect of this phenomenon, we consider the abundance
of kaons in a hadron gas. Given the ideal gas density of kaons,
$n_K(T)=m_K^2K_2(m_K/T)$, local strange\-ness conservation leads to
the suppressed form
\be
n_K(T,x) = n_K(T)\left\{{I_1(x) \over I_0(x)}\right\},
\label{GC}
\ee
where $x=V_0n_K(T)$ specifies the number of kaons inside the correlation
volume $V_0$; $I_0,~I_1$ and $K_2$ are the corresponding Bessel and
Hankel functions of imaginary argument. In the limit of high density or
large correlation volume, $x \to \infty$, $I_1(x)/I_0(x) \to 1$, and the
ideal gas abundances are correct. For low density or small correlation
volume, i.e., for $x \to 0$, $I_1(x)/I_0(x) \ll 1$, thus resulting in an
effective strangeness suppression as compared to the ideal gas abundance
of kaons (see Fig.\ \ref{local}).

\medskip

\vskip -1cm

\medskip

We are thus confronted with the puzzling question of how the transition
from suppressed to normal strangeness production occurs. What specifies
the correlation volume $V_0$? A rather
natural solution to this puzzle is that the abundance of strange hadrons
is already determined by the initial state partonic cluster size in the
transverse plane: the extension of this cluster specifies $V_0$. This
means that parton percolation and the corresponding sudden increase of
$V_0$ result in $I_1/I_0 \to 1$ and thus trigger the
transition to genuine ideal gas abundances of strange particles. More
detailed work on this is in progress; an interesting aspect is obviously
the resulting correspondence between charmonium suppression and
strangeness enhancement.

\subsection{Thermal or Statistical?}

This point is undoubtedly the most speculative of the present
considerations. At various stages of the high energy heavy ion program,
it has been asked whether a single collision event already produces a
thermal medium (`matter'), or whether each individual event is something
like one member of a Gibbs ensemble in thermodynamics, so that only an
average over many events results in a thermal pattern.

\medskip

\begin{figure}[h]
\centerline{\psfig{file=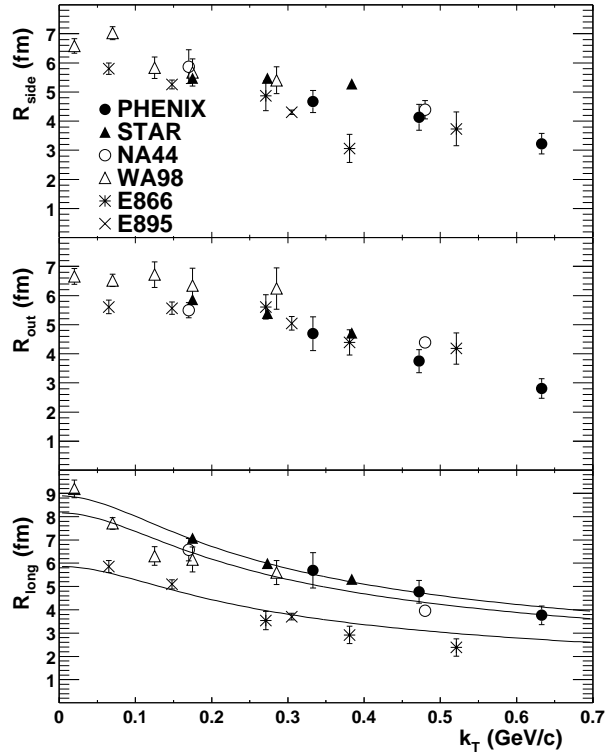,width=8cm}}
\caption{Source radii obtained from HBT interferometry in nuclear
collisions at different energies \cite{Adcox}.}
\label{hbt}
\end{figure}

If we take the extreme point of view that the observable phenomena are
determined fully determined by the initial state parton configuration
of the colliding nuclei, without any subsequent thermalization, then a
single collision does not lead to a thermodynamic medium. Everything is
specified by the given nuclear collision configuration, and there will
in particular not be any kind of `expanding and cooling matter'. This
suggests that the source size as determined by HBT interferometry
should essentially measure the initial nuclear collision configuration;
it should show no dependence on the collision energy and not lead to
any `medium life-time', with always $R_{\rm out} \simeq R_{\rm side}$.
Moreover, the transverse expansion velocity should also become constant
with increasing collison energy. These features are indeed observed, as
shown in Figs. \ref{pavel} and \ref{hbt}, contrary to all earlier
predictions.

\medskip

\medskip

Perhaps there exist models which nevertheless can reproduce the entire
set of present observations in thermal terms. However, it does seem
worthwhile to pursue further the possibility that initial state
conditions alone already determine them rather naturally, and to search
for some crucial additional observable which would unambiguously
indicate event-by-event thermalization.

\vskip 1cm

\noindent{\Large \bf Acknowledgements}

\bigskip

\noindent
It is a pleasure to thank S.\ Digal, S.\ Fortunato, F.\ Karsch, C.\ 
Louren{\c c}o, M.\ Nardi,
P.\ Petreczky and K.\ Redlich for many stimulating discussions
and helpful comments.


\vskip 1cm

\begin{thebibliography}{99}

\bibitem{KKG} J.\ R.\ Ellis and K.\ Geiger, \PL B404 (1997) 230.

\bibitem{LEP} For a recent survey, see P.\ Abreu, hep-ph/0111395.

\bibitem{stauffer} D.\ Stauffer and A.\ Aharony, {\sl Introduction to
Percolation Theory}, Taylor and Francis, London 1994.

\bibitem{GRV} M.\ Gl\"uck, E.\ Reya and A.\ Vogt, \ZP C 67 (1995) 433;\\
\EP C 5 (1998) 461.

\bibitem{isi}See e.\ g., M.\ S.\ Isichenko, Rev.\ Mod.\ Phys.\ 64 (1992) 961.

\bibitem{attig} H.\ Satz, Comput.\ Phys.\ Commun.\ 147 (2002) 46.

\bibitem{Torino} M.\ Nardi and H.\ Satz, \PL B 442 (1998) 14;\\ 
H.\ Satz, \NP A 661 (1999) 104c.

\bibitem{DFPS} S.\ Digal et al., \PL B 549 (2002) 101.

\bibitem{K-N} D.\ Kharzeev and M.\ Nardi, \PL B 507 (2001) 121.

\bibitem{KLNS} D.\ Kharzeev et al., \ZP C 74 (1997) 307

\bibitem{CGC} L.\ McLerran and R.\ Venugopalan, \PR D 49 (1994) 2233 and 
3352;\\
L.\ McLerran, Lect.\ Notes Phys.\ 583 (2002) 291.

\bibitem{NA50} M.\ C.\ Abreu et al.\ (NA50), \PL B 477 (2000) 28 and
\PL 521 (2001) 195

\bibitem{B-O} J.-P. Blaizot, M.\ Dinh and J.-Y.\ Ollitrault, 
\PRL 85 (2000) 4012.

\bibitem{BC} F. Becattini et al., Phys. Rev. C64 (2001) 024901.

\bibitem{Pavel} Compilation by P. Danielewicz, Nucl. Phys. A685 (2001) 368.

\bibitem{KR} R.\ Hagedorn and K.\ Redlich, \ZP C 27 (1985) 541;\\
J.\ Cleymans et al., \PR C 59 (1999) 1669.

\bibitem{Adcox} Compiled in K. Adcox et al. (PHENIX), Phys. Rev. Lett. 88
(2002) 192302.


\end{thebibliography}
\end{document}